\documentclass[fleqn,usenatbib,useAMS]{mnras}
\usepackage{natbib}
\usepackage{float} 
\usepackage{graphicx}	
\usepackage{amsmath}	
\usepackage{amssymb}	
\usepackage{multicol}        
\usepackage{bm}		
\usepackage{pdflscape}	
\usepackage{multirow}
\usepackage{comment}
\usepackage[normalem]{ulem} 
\usepackage{xcolor} 

\usepackage[T1]{fontenc}
\usepackage{ae,aecompl}
\usepackage{graphicx}
\usepackage{booktabs}
\usepackage{newtxtext,newtxmath}
\usepackage{natbib}

\title[Guwenbo]{Optical turbulence in the atmospheric surface layer at the Pamir Plateau Muztagh-ata site}

\author[Gu Wenbo et al.]{
	\large
	Wenbo Gu$^{1,2}$, Ali Esamdin$^{1,2}$\thanks{E-mail: aliyi@xao.ac.cn}, Chunhai Bai$^{1}$, Xuan Zhang$^{1}$, Guojie Feng$^{1}$,  Guangxin Pu$^{1}$, Letian Wang$^{1}$, {Gaowen Sun}$^{1}$ \\
	\large
	\textnormal{ Haozhi Wang$^{1,2}$, Lixian Shen}$^{1,2}$ \\
	$^{1}$Xinjiang Astronomical Observatory, Chinese Academy of Sciences, Urumqi 830011, China; guwenbo@xao.ac.cn , aliyi@xao.ac.cn\\
	$^{2}$University of Chinese Academy of Sciences, Beijing 100049, China
}

\date{Accepted 2024 October 21. Received 2024 October 21; in original form 2024 May 23 }
\pubyear{2024}

\begin{document}

\label{firstpage}
\pagerange{\pageref{firstpage}--\pageref{lastpage}}
\maketitle
\begin{abstract}
In this paper, we conducted a detailed analysis of optical turbulence in the Atmospheric Surface Layer (ASL) at Muztagh-ata site during on-site testing. We utilized ultrasonic anemometers positioned on a 30-meter tower to collect and process data at five height levels, obtaining data from October 1, 2021 to the present. We investigated the behavior of optical turbulence parameters (\(C_n^2\) and seeing \(\varepsilon\)) in the ASL. Nighttime \(C_n^2\) primarily fluctuated in the range of \(10^{-16}\) to \(10^{-14}\), exhibiting an exponential decrease with height. During the day, it showed a \(h^{-0.82}\) dependency, while at night, it displayed a \(h^{-0.48}\) dependency. Additionally, we presented the distribution of seeing across different layers within the ASL, showing a gradual decrease with increasing height, with a median seeing of 0.24 arcseconds at nighttime and 0.48 arcseconds at daytime between 6-30m. 
We investigated the relationship between surface temperature inversion, seeing in the ASL, and wind speed at the site. Our results show that under temperature inversion conditions, seeing significantly improves and is often accompanied by low to moderate wind speeds, while high wind speeds are usually associated with poorer seeing.
Preliminary calculations and observational results, combined with the high altitude and unique geographical location, suggest that Muztagh-ata site has the potential to be an outstanding optical astronomical observatory in the western plateau of china.
\end{abstract}

\begin{keywords}
Optical turbulence, atmospheric effects, site testing
\end{keywords}



\newpage
\section{Introduction}
Muztagh-ata site is located in the eastern Pamir Plateau region in the southwestern part of Xinjiang, China. The geographic coordinates of the main monitoring area are about $38^\circ 21'$ north latitude and $74^\circ 54'$ east longitude, with an altitude of about 4500 meters.
The monitoring of astronomical observation conditions at the site officially began in January 2017, the previous on-site testing activities have analyzed Astronomical Meteorology(Xu et al. \citeyear{xu2020a}\href{URL}{\textcolor{blue}{}}), Nocturnal Sky Brightness(Xu et al. \citeyear{xu2020a}\href{URL}{\textcolor{blue}{}}), Cloud Cover(Xu et al. \citeyear{xu2023}), Seeing(Xu et al. \citeyear{xu2020b}\href{URL}{\textcolor{blue}{}},\citeyear{xu2020c}\href{URL}{\textcolor{blue}{}}) and Atmospheric Water Vapor Content(Xu et al. \citeyear{xu2022}). These studies demonstrate the exceptional optical observing conditions at the site, highlighting its advantages for precise optical astronomical research. With the upcoming establishment of a 1.93-meter diameter telescopes at this location, it is poised to become a significant optical/infrared astronomical observatory in western China.
At the site, a 30-meter tower was established at an elevation of 4488m in September 2021 to begin data collection. A 6-meter DIMM (Differential Image Motion Monitor) tower and a 10-meter DIMM tower were successively established at elevations of 4510 meters and 4504 meters. In previous research, a median seeing of 0.82 arcseconds was recorded at the site. During the winter season, the median seeing can reach 0.60 arcseconds (Xu et al. \citeyear{xu2020b}\href{URL}{\textcolor{blue}{}}). The seeing conditions at the site reflect a lower impact of optical turbulence on optical astronomical observations, thereby favoring high-precision and high-resolution optical astronomical observations.


\begin{figure}
	\includegraphics[width=\columnwidth]{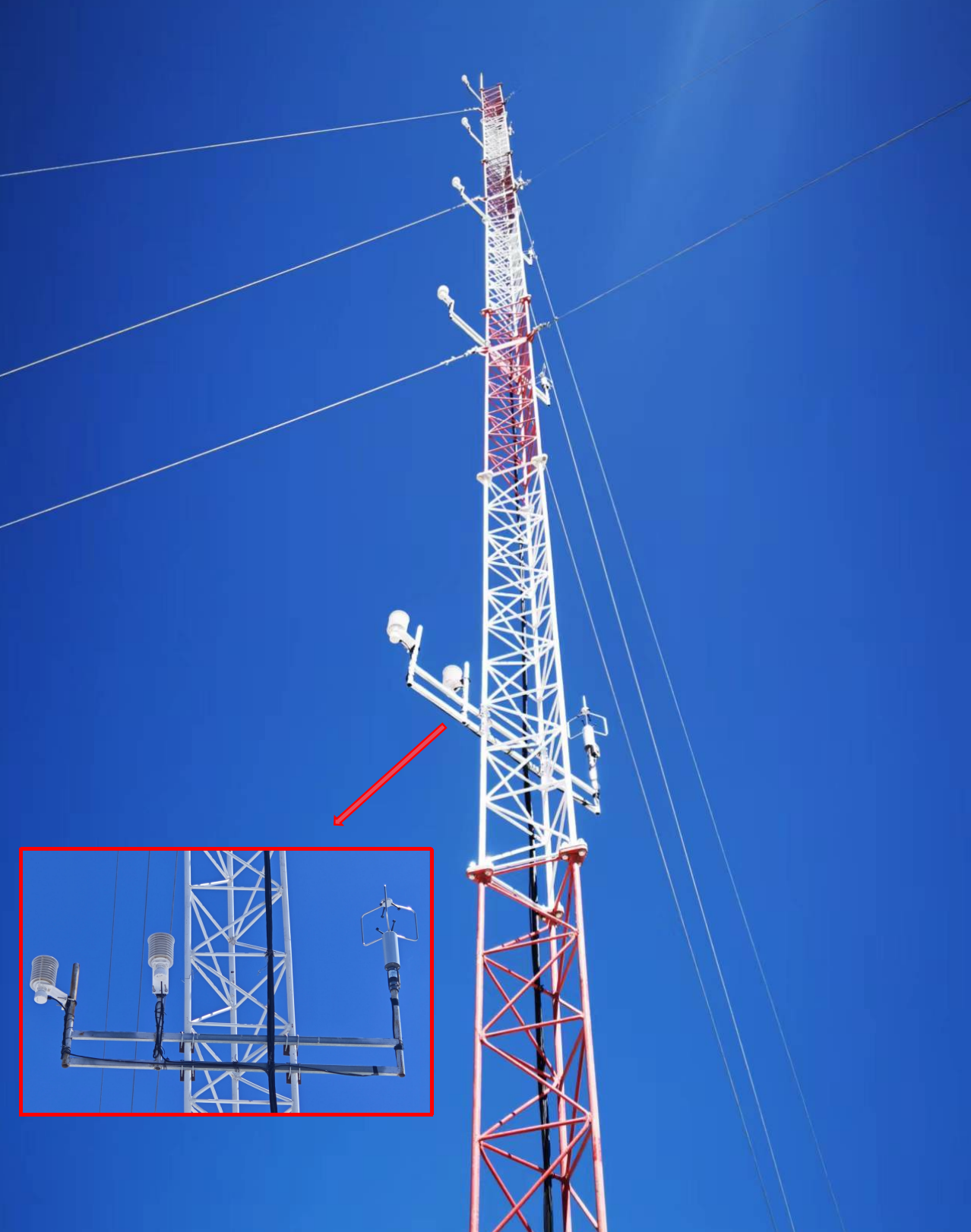}
	\caption{A 30-meter tower at Muztagh-ata site is equipped with temperature, humidity, and pressure sensors, as well as five layers of ultrasonic anemometers.}
	\label{fig:30m}
\end{figure}

\begin{table*}
	\centering
	\caption{The installation positions, precision, and Output frequency of each instrument.}
	\label{tab:Sensor Performance and Output Frequency}
	\begin{tabular}{|l|l|l|l|}
		\hline
		Sensor & Main Performance Parameters & Output frequency & Sensor on 30-meter tower \\
		\hline
		Young 81000 & Wind speed resolution: 0.01 m/s & 20 Hz & 6m,12m,18m,24m,30m\\
		            & Wind direction resolution: 0.1 °&   &  \\
		            & Sonic temperature resolution: 0.01\textdegree{}C&   &  \\
		Young 41342 & Temperature resolution: 0.1\textdegree{}C & 1/30 Hz & 2m,6m,12m,18m,24m,30m\\
		Young 61302V & Atmospheric pressure resolution: 0.2 hPa & 1/30 Hz & 2m\\
		Young 41382VC & Relative humidity resolution: 0.1\% RH & 1/30 Hz & 2m\\
		\hline
	\end{tabular}
\end{table*}
Turbulence is a highly irregular and chaotic macroscopic fluid motion, when light propagates through the atmosphere, random fluctuations in the refractive index caused by turbulent eddies lead to a series of optical turbulence effects, including wavefront distortion, beam wandering, and image scintillation that have influenced  the performance of optoelectronic systems, including astronomical observations, atmospheric optical communications, and optical remote sensing (Tatarskii \citeyear{1978Tatarskii},Andrews \& Phillips \citeyear{1999Andrews},Cui et al. \citeyear{2013cui},Zhou et al. \citeyear{2010zhou}).
The optical turbulence that affects visible light and near-infrared wavelengths is mainly caused by uneven distribution of temperature, humidity, and wind speed. Its intensity is measured by the refractive index structure constant ($C_n^2$).
Seeing($\varepsilon$) is a parameter that describes the degree of blurring of stellar images caused by atmospheric turbulence. It quantifies the angular range of phase disturbances experienced by the light waves. 
Optical turbulence displays a stratified distribution with altitude, and the disparity in turbulence conditions between layers is evident(Vinnichenko \citeyear{vinnichenko2013turbulence}; Ahrens \& Henson \citeyear{ahrens2015meteorology}).
The atmospheric surface layer(ASL) is the lowest part of the turbulent atmospheric boundary layer, extending several tens of meters above the ground(Stull \citeyear{stull1988}).  Typically, intense turbulence is primarily concentrated in the boundary layer, with optical turbulence in the vicinity of ASL playing a substantial role in the overall seeing conditions.(Abahamid et al. \citeyear{abahamid2004};Trinquet et al. \citeyear{trinquet2008};Agabi et al. \citeyear{agabi2006}). Hence, accurately measuring the spatiotemporal distribution patterns of optical turbulence within the ASL at Muztagh-ata site and assessing its impact on optical imaging hold significant importance. This process can provide crucial design guidelines for the construction of large optical telescopes at this location and further optimize adaptive optics (AO) systems.

In the study of optical turbulence within the ASL, various instruments can be used, each with its own characteristics and applications, such as Sonic Detection and Ranging (SODAR)(Lawrence et al. \citeyear{2004Natur.431..278L}; Petenko et al. \citeyear{2014A&A...568A..44P}; Qiang et al. \citeyear{2017Qiang}), Lunar Scintillometer (Lusci) (Hickson et al. \citeyear{2004Hickson}; Osborn \citeyear{osborn2010profiling}), Slope Detection and Ranging (SLODAR) (Wilson \citeyear{2002MNRAS}; Butterley et al. \citeyear{2006MNRASBUt}), Scintillation Detection and Ranging (SCIDAR) (Osborn et al. \citeyear{2010Osborn}), microthermal sensors (Azouit \& Vernin. \citeyear{2005Azouit}; Trinquet et al. \citeyear{2008PASP..120..203T}), and ultrasonic anemometers.
The ultrasonic anemometer, with its low energy consumption and strong anti-interference capability has been widely used for ASL meteorological parameter measurements.(Aristidi et al. \citeyear{2015Aristidi}; Qian et al. \citeyear{2022MNRASQ}). Kaimal (\citeyear{1979dmuf}) first discussed its application in atmospheric optical turbulence.
The principle behind the ultrasonic anemometer involves the measurement of ultrasonic virtual temperature and the three-dimensional wind field at a rapid pace, subsequently, the acquired da\label{key}ta is processed to estimate the optical turbulence parameters.
For this purpose, five ultrasonic anemometers have been installed on a 30-meter tower at Muztagh-ata site, at heights of 6, 12, 18, 24, and 30 meters above the ground. The measurements are conducted based on the Monin-Obukhov similarity theory(Obukhov \citeyear{Obukhov1971}) and the Wyngaard semi-empirical relations(Wyngaard et al. \citeyear{1971JOSA...61.1646W}). Figure~\ref{fig:30m} depicts the layout of the 30-meter tower at the site and shows the placement of the ultrasonic anemometers. Analysis on temporal and vertical distributions of optical turbulence parameters within ASL relies on the meteorological parameters obtained from our on-site measurements. By examining these distributions, we aim to gain a deeper understanding of the characteristics of optical turbulence at different heights within the ASL.

This study mainly focuses on the  optical turbulence in the ASL at Muztagh-ata site. The main results include the temporal and spatial distribution of the optical turbulence intensity \(C_n^2\) and the corresponding seeing statistics from October 2021 onwards, spanning a year. In section 2, we described the on-site measurement methods used to monitor optical turbulence at Muzagh-ata site. Theoretical framework and data analysis for these measurements are provided in Section 3. The measurement results for optical turbulence parameters in the ASL are presented in Section 4. The main findings are presented in Section 5.

\section{Measurements}

At Muztagh-ata site, we have erected a 30-meter tower featuring a gradient arrangement of instruments for monitoring atmospheric parameters. The main instruments installed include three-dimensional ultrasonic anemometers (Young 81000), temperature sensors (Young 41342), high-precision pressure sensors (Young 61302V), and relative humidity sensors (Young 41382VC). The primary sensor models and specifications are listed in Table~\ref{tab:Sensor Performance and Output Frequency}. This stratified arrangement enables high-resolution measurements of three-dimensional wind fields and temperatures at multiple elevations.

The ultrasonic anemometer (Young 81000) is utilized for measuring three-dimensional wind speed and direction, as well as sonic temperature, by analyzing the transit time of ultrasonic acoustic signals. Sonic temperature is derived from the speed of sound, which is adjusted for crosswind effects. This anemometer offers an adjustable output frequency ranging from 4 to 32 Hz, with a wind speed measurement range of 0 to 40 m/s and a resolution of 0.01 m/s. The accuracy of wind speed measurements is ±1\% rms within the 0 to 30 m/s range, and ±3\% rms within the 30 to 40 m/s range. The wind direction is resolved with a precision of 0.1 degrees and an accuracy of ±2° for wind speeds between 1 and 30 m/s, and ±5° for wind speeds between 30 and 40 m/s. Sonic temperature measurements span a range of -50 to +50 °C, with a resolution of 0.01 degrees Celsius. Its accuracy is ±2°C, even when wind speeds are below 30 m/s.
The relative humidity and barometric pressure sensors acquire data with a sampling interval of 30 seconds, facilitating high-precision measurements of relative humidity and atmospheric pressure.

\begin{figure}
	\includegraphics[width=\columnwidth]{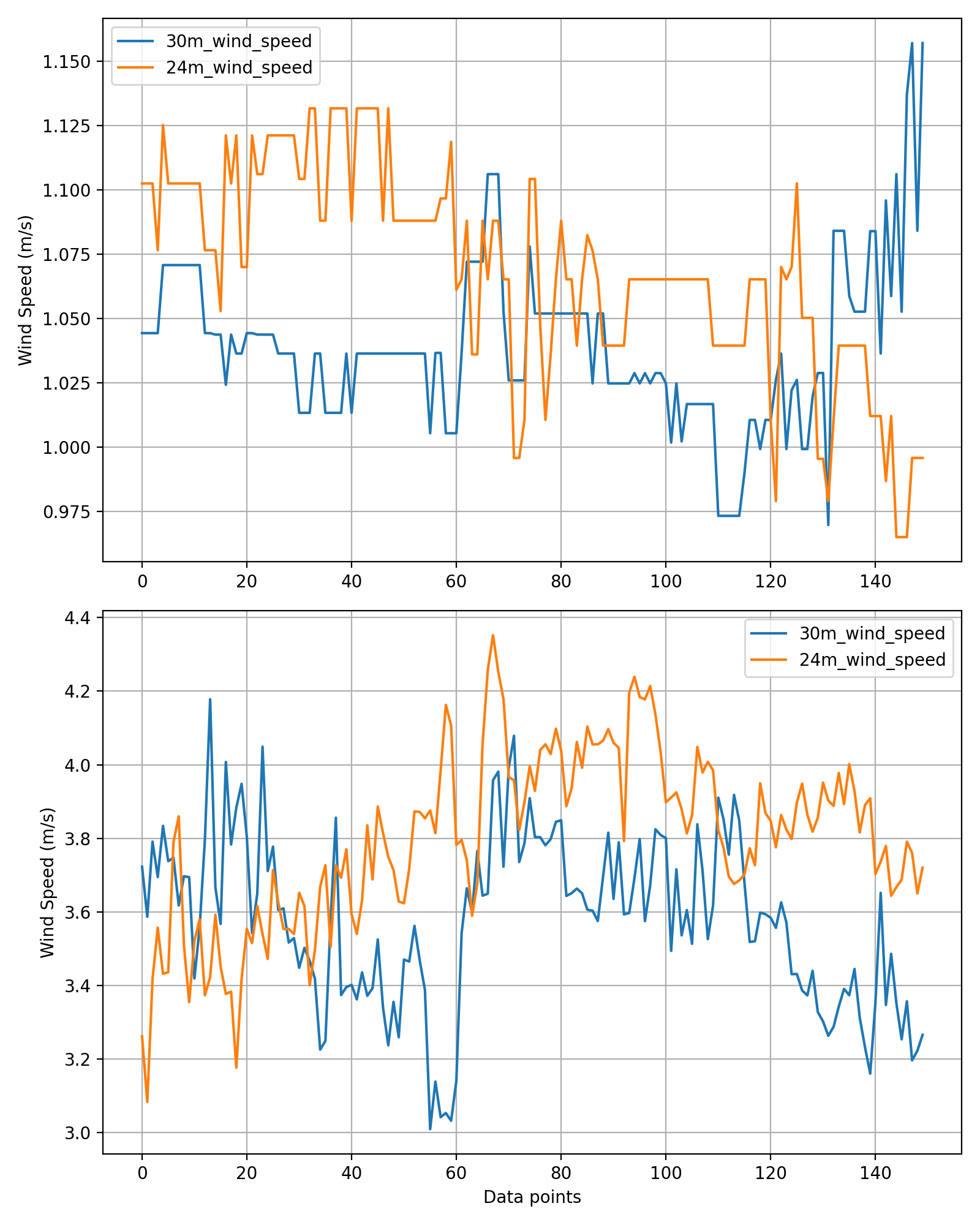}
	\caption{ The comparison between normal and anomalous data is depicted, with the upper section showing anomalous data and the lower section showing normal data. The x-axis displays consecutive data numbers, while the y-axis represents the magnitude of wind speeds in meters per second (m/s). }
	\label{fig:combined_wind_speed_plots.png}
\end{figure}

The system commenced observations on October 1, 2021 and data collection has been ongoing since then, resulting in approximately 100MB of data generated daily. The dataset utilized in this study comprises 335 days of available data collected from October 1, 2021 to September 30, 2022.

During preliminary analysis of the ultrasonic anemometer data, instances of unreasonable data were identified which are characterized by significant duplication.
In Figure \ref{fig:combined_wind_speed_plots.png}, we present the noticeable duplicate data observed over a certain period. The top half of Figure \ref{fig:combined_wind_speed_plots.png} illustrates the anomalies encountered during the data collection with the ultrasonic anemometer, revealing a substantial amount of repeated data. In contrast, the bottom half of the figure displays the data under normal operational conditions. This visualization offers a comparative insight between the abnormal data points and the normal ones.
To address the notable impact of duplicated data on our subsequent work, we implemented a filter to handle such repetitions. Additionally, the reasons behind the generation of such data by the ultrasonic anemometer were investigated. After a thorough examination of all data, there might be two potential causes of anomalous data: 
The first reason is the frequent occurrence of abnormal data collection by the sonic anemometer under rainy, snowy, or frosty conditions. The second reason is the apparent insensitivity of the sonic anemometer to extremely low wind speeds and calm conditions in high-altitude environments, resulting in prolonged consecutive duplicate values at wind speeds below 1 m/s. 
To mitigate this issue, we segmented all data into 15-minute intervals for detailed inspection, identifying and eliminating duplicate data within each time block. Through comparisons across multiple datasets, we assessed the impact of anomalous data on temperature structure functions. Following careful evaluation, a suitable threshold was established: if the proportion of anomalous data within a 15-minute interval does not exceed 40\%, the data quality remains satisfactory after excluding these outliers. Applying this filtering criterion to the entire dataset, we assessed all 15-minute time intervals, revealing that the accumulated proportion of periods with anomalous data exceeding 40\% is 10\%. This rigorous data filtering process ensures the reliability of our subsequent calculations and analyses, ultimately establishing a highly reliable dataset from the ultrasonic anemometer.

\section{Theory and Data Processing}
\subsection{Theory}
The optical turbulence affecting the fluctuations in refractive index in the visible and near-infrared spectral bands is primarily induced by temperature variations. Temperature structure function $D_{T}(\vec{\rho})$ is one of the statistical quantities describing the temperature fluctuations between two points separated by a distance $\rho$, which is expressed as:
\begin{equation}
D_{T}(\vec{\rho}) = \left\langle (T(\vec{r}) - T(\vec{r} + \vec{\rho}))^2 \right\rangle
\end{equation}

Where $T$ is the temperature, $\vec{r}$ is the position of one point, and $\langle \rangle$ denotes ensemble average. 
According to the assumption of locally homogeneous and isotropic turbulence by Kolmogorov and under the assumption of fully developed turbulence, the temperature structure function ($D_{T}(\vec{\rho})$) satisfies the inertial range scales ($l_0$) and outer scales ($L_0$) (Kolmogorov \citeyear{1941DoSSR..30..301K}; Tatarskii \citeyear{1961Tatarskii}, \citeyear{1978Tatarskii}):
\begin{equation}
	D_{T}(\vec{\rho})=C_{T}^{2} \rho^{2/3}
\end{equation}

where \(C_T^2\) is the temperature structure constant, and \(\rho\) is the distance between two measurement points. The range [\(l_0-L_0\)] is commonly referred to as the inertial subrange, within which energy is primarily transferred through interactions between vortices rather than direct dissipation, turbulent vortices at these scales are the main contributors to beam path disturbance and phase distortions (Lukin \citeyear{2005SPILukin}; Ziad \citeyear{2016Ziad} ). 

For the actual single-point temperature measurements obtained from the three-dimensional ultrasonic anemometer, the temperature structure function is derived based on the Taylor freezing hypothesis (Taylor \citeyear{1932Taylor}), using the temperature fluctuations at different heights, we have to transform $\rho$ as $\overline{v}\Delta t$. Where $\overline{v}$ is the average wind speed, \( \Delta t \) is the time interval between two measurements, and the temperature structure function is calculated as follows:
\begin{equation}
D_{T}(\overline{v}\Delta t) = \left\langle (T(t) - T(t + \Delta t))^2 \right\rangle
\end{equation}

According to Gladstone's law, the relationship between $C_{n}^{2}$ and $C_{T}^{2}$ is given by:
\begin{equation}
C_{n}^{2} =6.24 \times 10^{-9}C_{T}^{2}{P}^{2}{T}^{-4}
\end{equation}
Here, $P$ represents the current atmospheric pressure in hPa, and $T$ represents the temperature in Kelvin.

Seeing($\varepsilon$) is typically obtained by integrating $C_n^2$ along the altitude $z$ (Roddier \citeyear{1981PrOpt..19..281R}). The formula is as follows:
\begin{equation}
\epsilon=5.25 \lambda^{-1 / 5}\left[\int_{h_{0}}^{\infty} C_{n}^{2}(z) \mathrm{d} z\right]^{\frac{3}{5}}
\end{equation}
where $\lambda$ represents the wavelength, typically taken as 500 nm for the visible light spectrum, and $\varepsilon$ is in radian here.

\subsection{Data Processing}

The ultrasonic anemometer outputs 20 sets of data per second, including wind speed data in three directions: U, V, and W, where W represents the vertical direction. Additionally, the output includes $T_s$, which stands for the sonic temperature. In wet air, the sound speed c varies as a function of temperature and humidity. We can correct $T_s$ and obtain the true temperature using the following formula(Wang \& Wu \citeyear{2014wxq}; Qian et al. \citeyear{2022MNRASQ}):
\begin{equation}
T=\frac{T_{s}}{1+0.608 q}
\end{equation}
Where $q$ is the specific humidity.
According to Kolmogorov's homogeneous and isotropic turbulence theory, under fully developed turbulence, we can construct $C_{T}^{2} \rho^{\alpha}$ according to Equation 2 and perform a least squares fit within the inertial range on a log-log scale. If the fitting slope $\alpha$ deviates too much from $\frac{2}{3}$, it indicates that the function does not follow the power law (Aristidi et al. \citeyear{2015Aristidi}).

When constructing the refractive index structure function, the ensemble average over time must be calculated over a sufficiently long time interval \(\tau\) to ensure statistical significance but shorter than the characteristic time of evolution of \(C_{T}^2\). After some trials, we found \(\tau = 15\) minutes to be a suitable choice.
Therefore, we first divide the data into 15-minute time intervals, and examine the fitted slope \(\alpha\) in batches. If \(\alpha < 0.4\), we reject the 15-minute time interval, because it does not comply with the power law.

Using this method, we filtered the data and then calculated \(C_{T}^{2}\) using the following formula:
\begin{equation}
	C_{T}^{2}=\left\langle\frac{[T(t)-T(t+\delta t)]^{2}}{(v\delta t)^{\frac{2}{3}}}\right\rangle_{\tau}
\end{equation}
This method calculates a specific value of the structure function at the smallest possible time interval \(\delta t = 0.05\) seconds. This value is represented by the quantity \([T(t)-T(t+\delta t)]^2\). \(C_T^2\) is estimated through this quantity, where the time average is taken over \(\tau = 1\) minute (Tatarskii \citeyear{1961Tatarskii}; \citeyear{1978Tatarskii}; Aristidi et al. \citeyear{2015Aristidi}).
where \(v\) is the wind speed, \(v =\sqrt{u^2+v^2+w^2}\), and \(u\), \(v\), and \(w\) represent three orthogonal components of the wind wind direction. The structural constant of refractivity \(C_n^2\) can be derived from Equation 4, where \(T\) takes the average temperature over 1 minute, and \(P\) takes the average pressure over 1 minute.

\begin{figure}
	\centering
	\begin{minipage}{\columnwidth}
		\includegraphics[width=0.99\columnwidth]{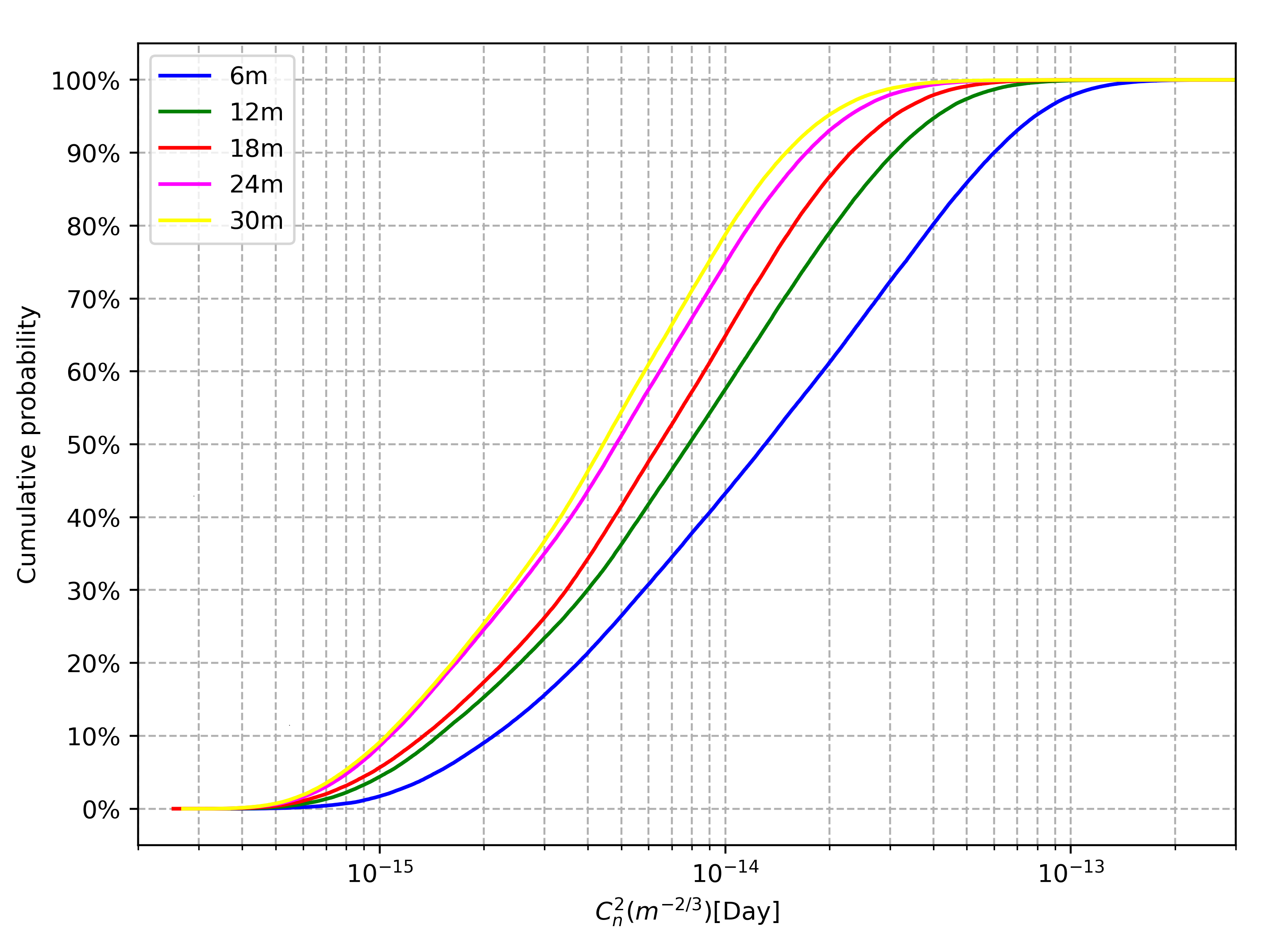}
		\vspace{-2mm} 
		\caption{The cumulative probability distribution of \( C_n^2 \) at five different heights during daytime at Muztagh-ata site.}
		\vspace{-1mm} 
		\label{fig:DayCP}
	\end{minipage}
	
	\begin{minipage}{\columnwidth}
		\includegraphics[width=0.99\columnwidth]{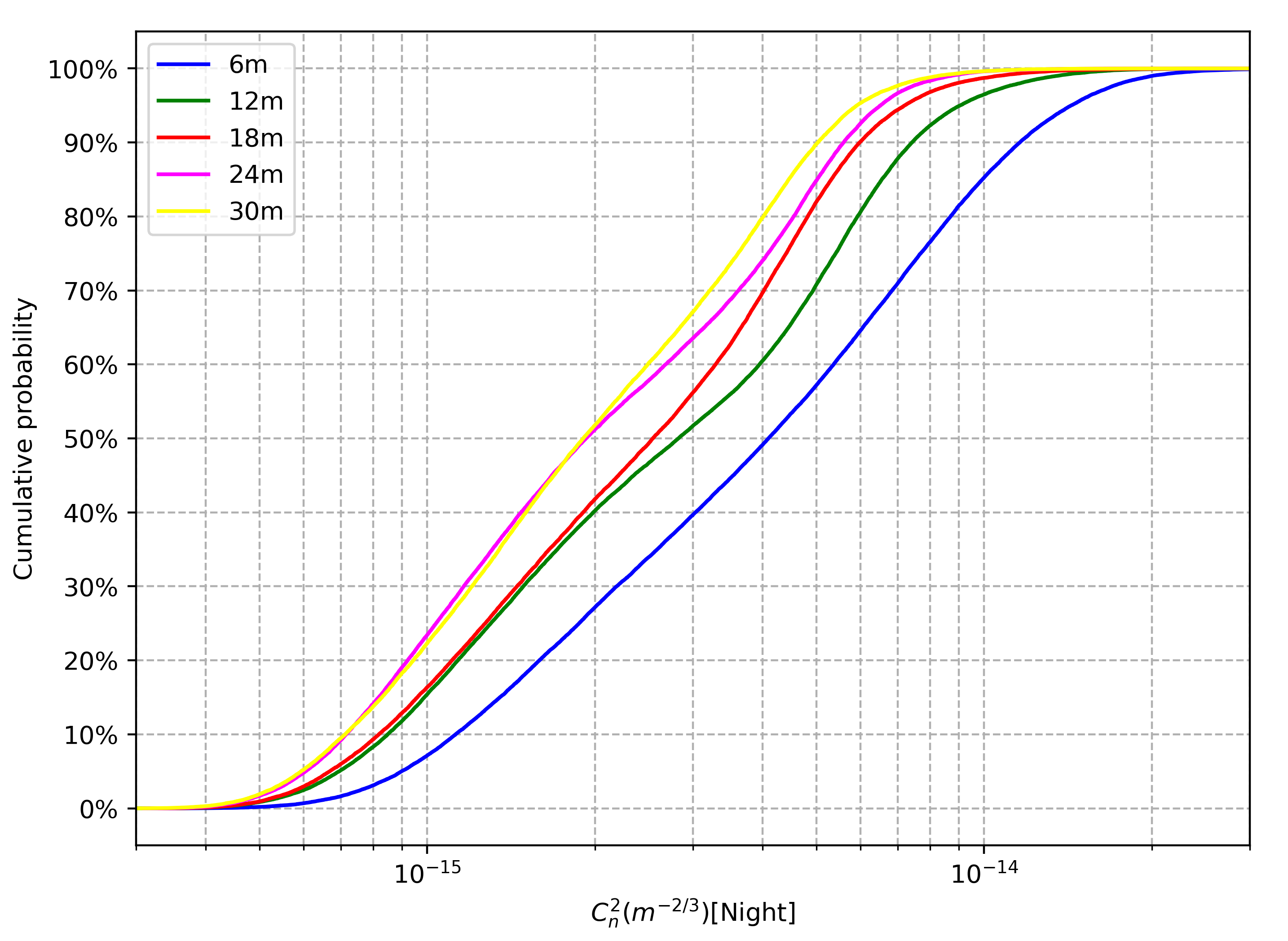}
		\vspace{-2mm} 
		\caption{The cumulative probability distribution of \( C_n^2 \) at five different heights during nighttime at Muztagh-ata site.}
		\vspace{-1mm} 
		\label{fig:NightCP}
	\end{minipage}
\end{figure}

\begin{figure}
	\centering
	\begin{minipage}{\columnwidth}
		\makebox[\columnwidth][c]{%
			\includegraphics[width=1.05\columnwidth]{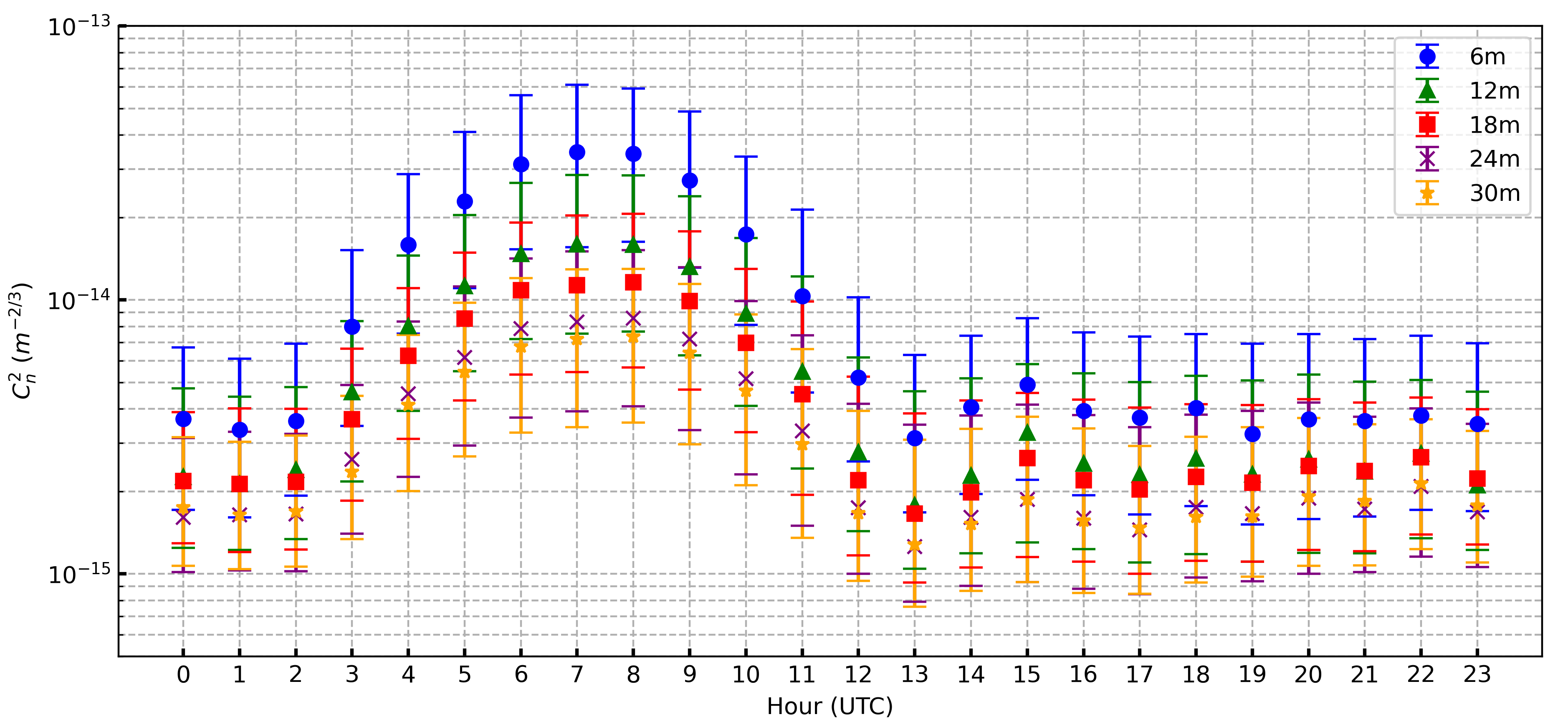}
		}
		\vspace{-3mm} 
		\caption{The diurnal variation of quartiles of \(C_n^2\) at heights of 6, 12, 18, 24, and 30 meters above the ground during the observation period.}
		\vspace{-1mm} 
		\label{fig:fig12}
	\end{minipage}
	
	\vspace{3mm} 
	
	\begin{minipage}{\columnwidth}
		\makebox[\columnwidth][c]{%
			\includegraphics[width=1.06\columnwidth]{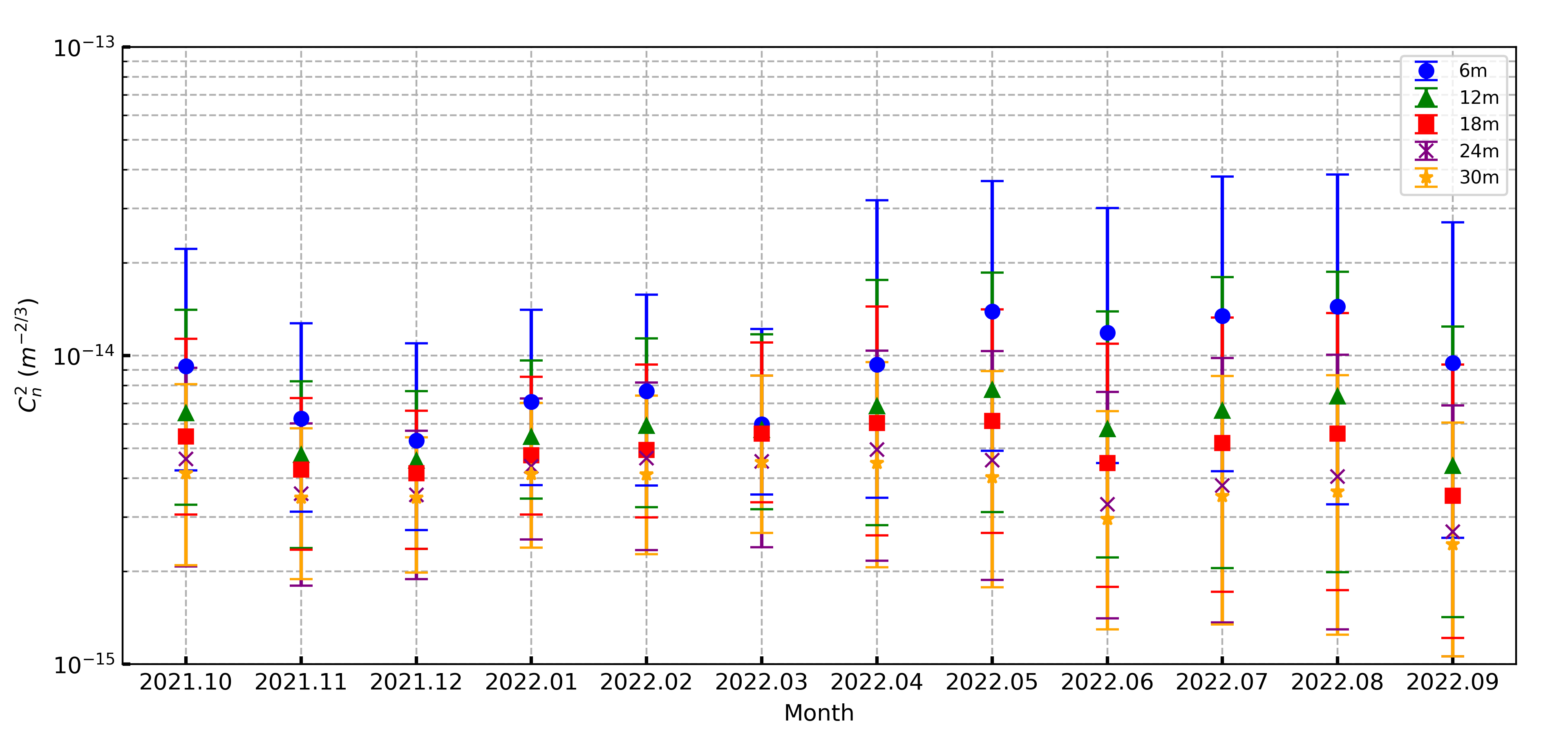}
		}
		\vspace{-3mm} 
		\caption{The monthly variation of quartiles of \(C_n^2\) at heights of 6, 12, 18, 24, and 30 meters above the ground during the observation period.}
		\vspace{-1mm} 
		\label{fig:fig13}
	\end{minipage}
\end{figure}

\section{results}
This study is based on Kolmogorov's theory of local isotropic turbulence. 
To ensure that our analysis is grounded in Kolmogorov's theoretical framework rather than being influenced by irregular turbulence behavior or non-standard turbulence evolution processes in the dataset, we filtered the data for each altitude, the filtered dataset conforms to theoretical expectations. Finally, we obtained a total of 979072 valid $C_{n}^{2}$ data points from five altitudes. In this dataset, the total number of daytime data points is 729749, while the nighttime data points amount to 249323. The dataset distribution is as follows: 20\% for spring, 30\% for summer, 26\% for autumn and 24\% for winter. All subsequent results are based on this dataset.
Additionally, we distinguish between day and night using twilight times. March, April, and May are considered spring; June, July, and August are considered summer; September, October, and November are considered autumn; and December, January, and February are considered winter. All times mentioned in this paper are in UTC. 

\begin{figure}
	\centering
	\includegraphics[width=\columnwidth]{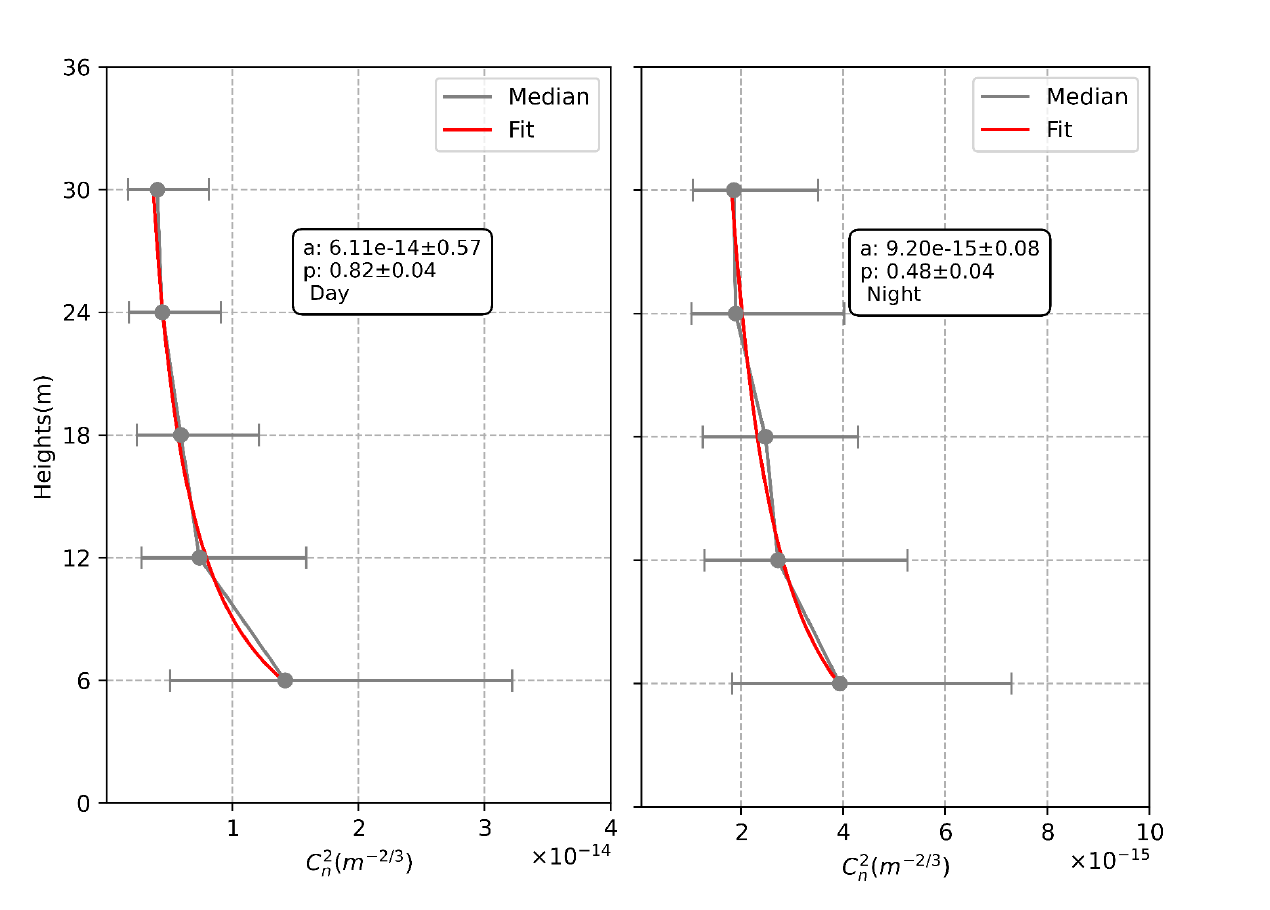}
	\vspace{-5mm} 
	\caption{The variation pattern of \( C_n^2 \) with respect to height at Muztagh-ata site, with daytime data on the left and nighttime data on the right. The red curve represents polynomial fitting, and data points represent corresponding medians. Error bars depict 50\% confidence intervals.}
	\vspace{-2mm} 
	\label{fig:season}
\end{figure}


\subsection{\(C_n^2\) in the ASL}

\begin{table}
	\caption{Seasonal median values and the 75th and 25th percentiles, with "Total" representing the data for the entire year. All values are in units of $10^{-15}$ m$^{-2/3}$. The definitions of the four seasons (summer, autumn, winter, and spring) are provided in the paper.}
	\label{tab:landscapesedffds}
	\resizebox{\columnwidth}{!}{
		\begin{tabular}{c|c|c|c|c|c|}\hline
			Height & \( C_{n}^{2} \) (Total)  &\( C_{n}^{2} \) (Spring)& \( C_{n}^{2} \) (Summer) & \( C_{n}^{2} \) (Autumn) & \( C_{n}^{2} \) (Winter) \\
			$[m]$  & $[10^{-15} m^{-2/3}]$ & $[10^{-15} m^{-2/3}]$  & $[10^{-15} m^{-2/3}]$   & $[10^{-15} m^{-2/3}]$ & $[10^{-15} m^{-2/3}]$\\ \hline
			6 & $9.4 \left[\begin{array}{c}26.2\\3.5\end{array}\right.$& $11.9 \left[\begin{array}{c}33.7\\4.0\end{array}\right.$   & $12.1 \left[\begin{array}{c}35.1\\2.9\end{array}\right.$    & $9.3 \left[\begin{array}{c}23.1\\3.7\end{array}\right.$       &  $7.2 \left[\begin{array}{c}15.5\\3.4\end{array}\right.$   \\\hline
			12  & $5.7  \left[\begin{array}{c}13.7\\2.2\end{array}\right.$& $7.0 \left[\begin{array}{c}17.4\\2.6\end{array}\right.$  & $5.8 \left[\begin{array}{c}16.5\\1.6\end{array}\right.$   &  $5.6 \left[\begin{array}{c}12.2\\2.2\end{array}\right.$   & $5.2\left[\begin{array}{c}9.7\\2.5\end{array}\right.$            \\ \hline
			18   & $4.7 \left[\begin{array}{c}10.8\\2.0\end{array}\right.$& $5.9 \left[\begin{array}{c}13.7\\2.5\end{array}\right.$ & $4.5 \left[\begin{array}{c}12.3\\1.4\end{array}\right.$  &  $4.6 \left[\begin{array}{c}9.6\\2.0\end{array}\right.$  &  $4.5 \left[\begin{array}{c}8.1\\2.3\end{array}\right.$            \\ \hline
			24    & $3.9 \left[\begin{array}{c}8.3\\1.5\end{array}\right.$& $4.6 \left[\begin{array}{c}10.2\\1.8\end{array}\right.$ & $3.3 \left[\begin{array}{c}9.0\\1.2\end{array}\right.$   & $3.7 \left[\begin{array}{c}7.5\\1.6\end{array}\right.$   &  $4.0 \left[\begin{array}{c}6.9\\1.9\end{array}\right.$    \\ \hline
			30   & $3.6 \left[\begin{array}{c}7.5\\1.5\end{array}\right.$& $4.2 \left[\begin{array}{c}9.1\\1.8\end{array}\right.$ & $3.0 \left[\begin{array}{c}7.8\\1.2\end{array}\right.$    & $3.4\left[\begin{array}{c}6.8\\1.6\end{array}\right.$     &  $3.7 \left[\begin{array}{c}6.5\\1.9\end{array}\right.$    \\ \hline
		\end{tabular}
	}
\end{table}
The method described in Section 3 was used to process data obtained from 30m tower to compute the \( C_{n}^{2} \) at various heights. Table ~\ref{tab:landscapesedffds} displays statistical data of $C_n^2$ obtained at each altitude (median and 50\% confidence interval).
As shown in it, optical turbulence near the surface layer is strongest during summer and weakest during winter, with optical turbulence intensity in summer approximately 2-3 times that of winter. The variation in it at different heights during summer is the most significant and correlates with surface thermal radiation.

Figures ~\ref{fig:DayCP} and ~\ref{fig:NightCP} show the cumulative probability distributions of \(C_{n}^{2}\) at heights of 6, 12, 18, 24, and 30 meters during both daytime and nighttime, respectively. During daytime, observations indicate that approximately 58\% of the time at heights of 12, 18, 24, and 30 meters have \(C_{n}^{2}\) values lower than 10\(^{-14}\) m\(^{-2/3}\), while at 6 meters, it's over 43\%. Almost 97\% of observations at 6 meters have \(C_{n}^{2}\) values lower than 10\(^{-13}\) m\(^{-2/3}\), and 100\% at other heights.
During nighttime, over 96\% of observations at heights of 12, 18, 24, and 30 meters have \(C_{n}^{2}\) values lower than 10\(^{-14}\) m\(^{-2/3}\), while at 6 meters, it's around 86\%. Almost all observations at each height have \(C_{n}^{2}\) values lower than 10\(^{-13}\) m\(^{-2/3}\).

\begin{figure}
	\centering
	\begin{minipage}{\columnwidth}
		\includegraphics[width=1.01\columnwidth]{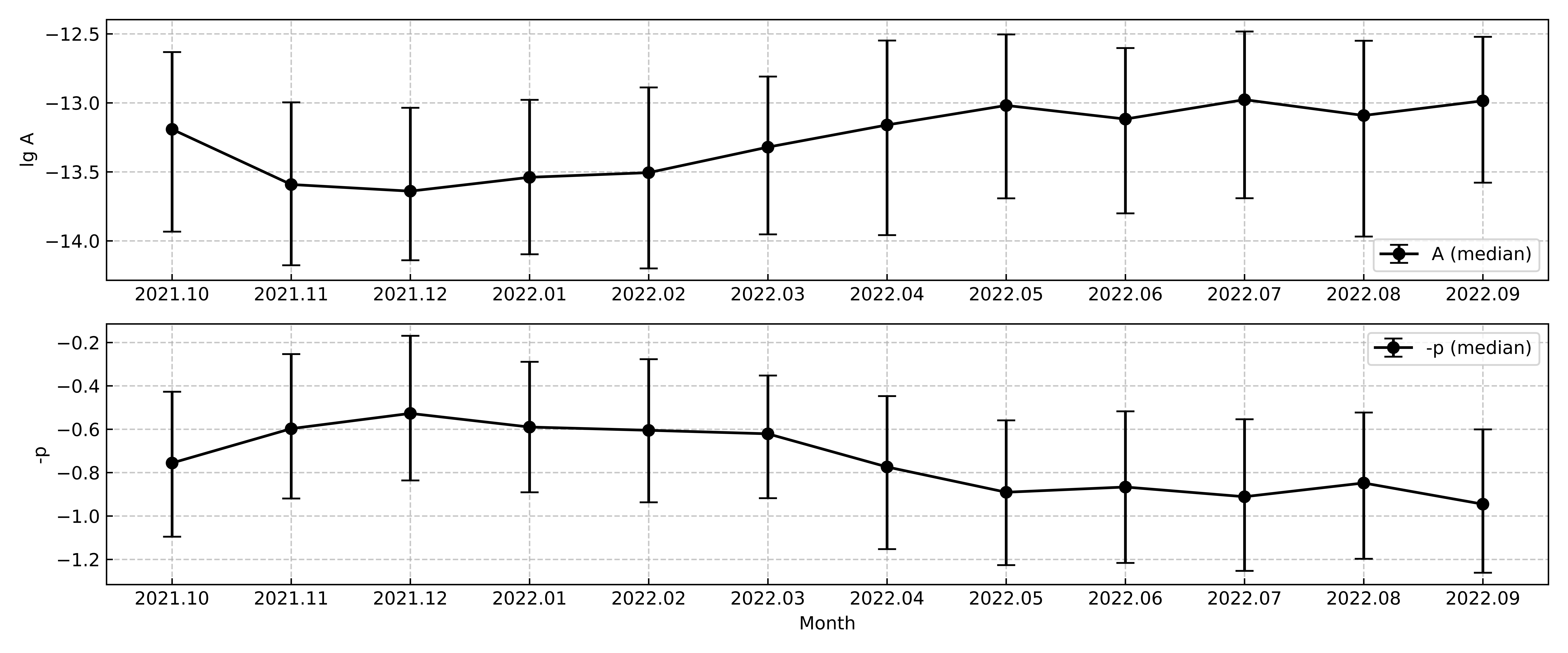}
		\vspace{-3mm} 
		\caption{The monthly variation patterns of the fitting parameters \(A\) and \(-p\) are presented. Error bars represent the 50\% confidence intervals.}
		\vspace{-1mm} 
		\label{fig:fig14}
	\end{minipage}
	
	\begin{minipage}{\columnwidth}
		\includegraphics[width=1.01\columnwidth]{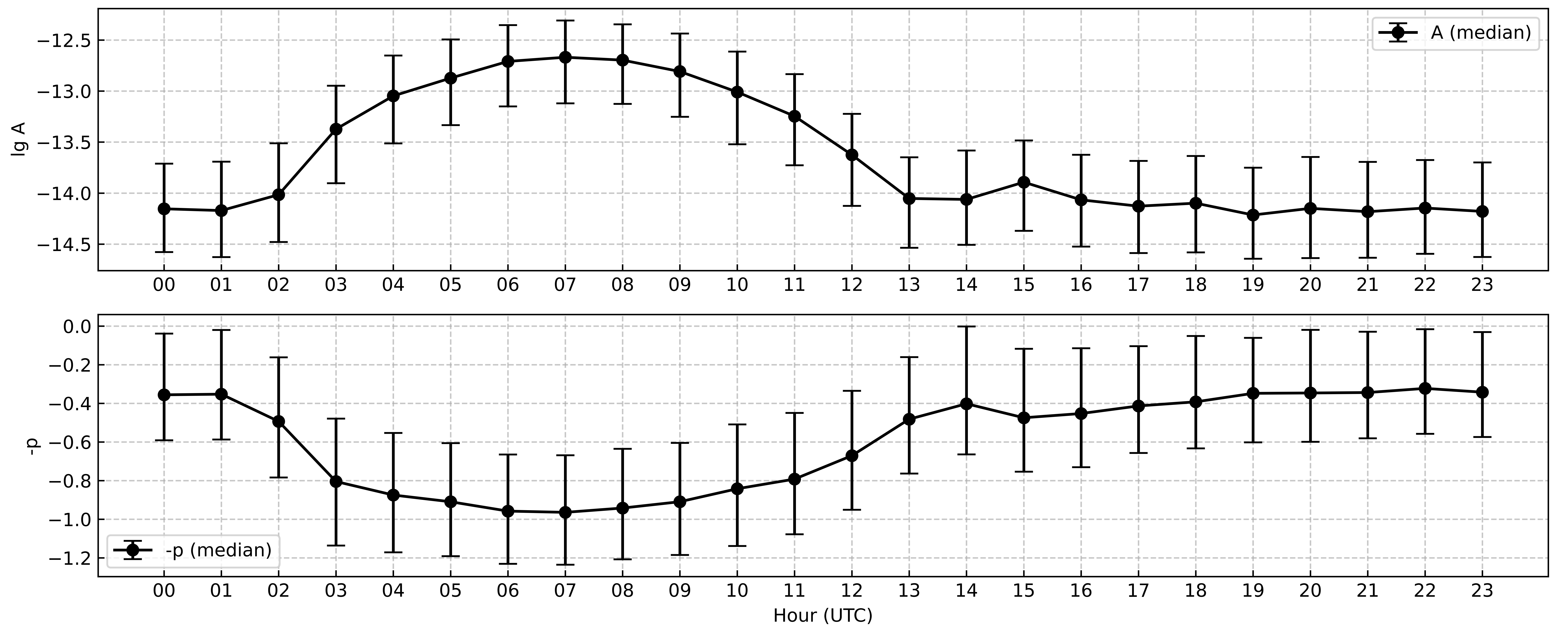}
		\vspace{-3mm} 
		\caption{The hourly variation patterns of the fitting parameters \(A\) and \(-p\) are shown. Error bars represent the 50\% confidence intervals.}
		\vspace{-1mm} 
		\label{fig:fig15}
	\end{minipage}
\end{figure}

Figure ~\ref{fig:fig12} illustrates the diurnal variation of \(C_n^2\) quartiles at heights of 6, 12, 18, 24, and 30 meters above the ground over the observation period. The diurnal variation of \(C_n^2\) at each height is essentially consistent, with it increasing around 11 a.m. and peaking around 3 p.m. Combined with lower value during the night, this variation is attributed to intense local turbulence caused by solar radiation heating.
Figure ~\ref{fig:fig13} presents the monthly variation of \(C_n^2\) quartiles at heights of 6, 12, 18, 24, and 30 meters above the ground. \(C_n^2\) is higher in summer compared to winter. It is noted that at Muztagh-ata site, \(C_n^2\) values primarily range between \(10^{-16}\) and \(10^{-13} \, \text{m}^{-2/3}\), with the maximum \(C_n^2\) value occurring at 6 meters above the ground.

Within the ASL, the \( \ln C_{n}^{2}(z) \) exhibits a trend of first-order polynomial variation with altitude. The vertical distribution of \( C_{n}^{2} \) primarily follows a negative exponential growth with height and is described as(Abahamid et al. \citeyear{abahamid2004}; Aristidi et al. \citeyear{2015Aristidi}):
\begin{equation}
	C_{n}^{2}(z) = A \cdot z^{-p}
\end{equation}
where \( z \) is the height, and \( A \) and \( p \) are fitting parameters.

To investigate the vertical distribution of optical turbulence intensity, we conducted polynomial fitting of the data collected at five equidistant heights using Equation 8. As illustrated in Figure~\ref{fig:season}, we present the vertical distribution of \(C_n^2\) in the ASL during daytime and nighttime at  Muztagh-ata site. The red curves represent polynomial fits, with data points indicating the corresponding medians. Error bars show the 25th and 75th percentiles of the data.
According to the data presented in Figure~\ref{fig:season}, the vertical distribution of \(C_n^2\) in the ASL at the site exhibits negative exponential decay, particularly noticeable during daytime. The descent rate between 6 and 12 meters consistently exceeds that at other heights. As altitude increases, the gradient decreases, suggesting that radiative cooling primarily affects the few meters above the ground. Daytime \(C_n^2\) values are significantly higher than nighttime values. The rate of decrease in \(C_n^2\) with increasing height below 18 meters during the day is notably higher than at night, possibly due to the inhibitory effect of the nocturnal near-surface temperature inversion layer on optical turbulence.

By analyzing statistical data of median \(C_n^2\) values obtained at each height during both daytime and nighttime, we can describe the fitted functions for \(C_n^2\) in the ASL at Muztagh-ata site as follows:
\begin{equation}
C_{n Day}^2 = (6.11 \pm 0.57) \times 10^{-14} \cdot z^{-0.82 \pm 0.04}
\end{equation}
\begin{equation}
C_{n Night}^2 = (0.92 \pm 0.08) \times 10^{-14} \cdot z^{-0.48 \pm 0.04}
\end{equation}

In this study, Equations 9 and 10 reveal a negative exponential dependence of \(C_n^2\) on height (\(z\)) at Muztagh-ata site. The fitted functions for both daytime and nighttime exhibit relatively lower \(p\) values, differing significantly from the \(C_n^2\) dependence on \(z^{-1.27}\) observed at Dome C on the Antarctic plateau (Aristidi et al. \citeyear{2015Aristidi}), as well as the \(C_n^2\) dependence on \(z^{-1.1}\) observed at Ali (Qian et al. \citeyear{2022MNRASQ}). This difference could be attributed to terrain or wind speed profiles.

Figure ~\ref{fig:fig14} illustrates the monthly variation of the median values of \(A\) and \(-p\) in the ASL at the site, along with quartiles representing errors. In summer, \(A\) is larger, and \(-p\) is smaller. 
Figure ~\ref{fig:fig15} illustrates the diurnal variation of the median values of \(A\) and \(-p\) in the ASL at the site, along with quartiles representing errors. \(A\) remains smaller during the night, slightly increasing after noon. On the other hand, \(-p\) is larger during the night and decreases during the day, reflecting the energy exchange between surface and air due to radiative heating and cooling. \(A\) values are notably smaller during the night, especially around astronomical dawn. Throughout the entire night, \(-p\) predominantly hovers around -0.5; In the afternoon, \(-p\) drops to around -0.9.

\subsection{ Seeing }

Seeing  is employed to characterize the level of star image degradation caused by atmospheric turbulence. This parameter holds significance in optical observations as it constitutes one of the key factors under consideration. Astronomical seeing can be quantified using Equation 5, which involves an integration of \(C_{n}^{2}\).
Starting from the first measurement point at 6 meters with the ultrasonic anemometer, we can compute seeing between each pair of heights. We used \(C_n^2\) data from five heights to calculate seeing, with the data sample for seeing calculation comprising approximately 80\% of the total \(C_n^2\) dataset.
The statistical results of seeing between various layers are presented in Table~\ref{tab:seeing111}. It can be found that seeing decreases with increasing altitude, with the highest seeing occurring between 6-12m.

\begin{table}
	\centering
	\caption{The statistical results of seeing within the layers of 6-12 meters, 12-18 meters, 18-24 meters, and 24-30 meters.}
	\label{tab:seeing111}
	\begin{tabular}{cccc}
		\toprule
		\textbf{height} & \multicolumn{3}{c}{\textbf{seeing (arcsec)}} \\
		\cmidrule(lr){2-4}
		& \textbf{25\%$\leq$} & \textbf{median} & \textbf{75\%$\leq$} \\
		\midrule
		6m-12m & 0.12 & 0.22 & 0.37 \\
		12m-18m & 0.10 & 0.17 & 0.28 \\
		18m-24m & 0.08 & 0.15 & 0.23 \\
		24m-30m & 0.08 & 0.13 & 0.20 \\
		\bottomrule
	\end{tabular}
\end{table}

\begin{figure}
	\centering
	\includegraphics[width=\columnwidth]{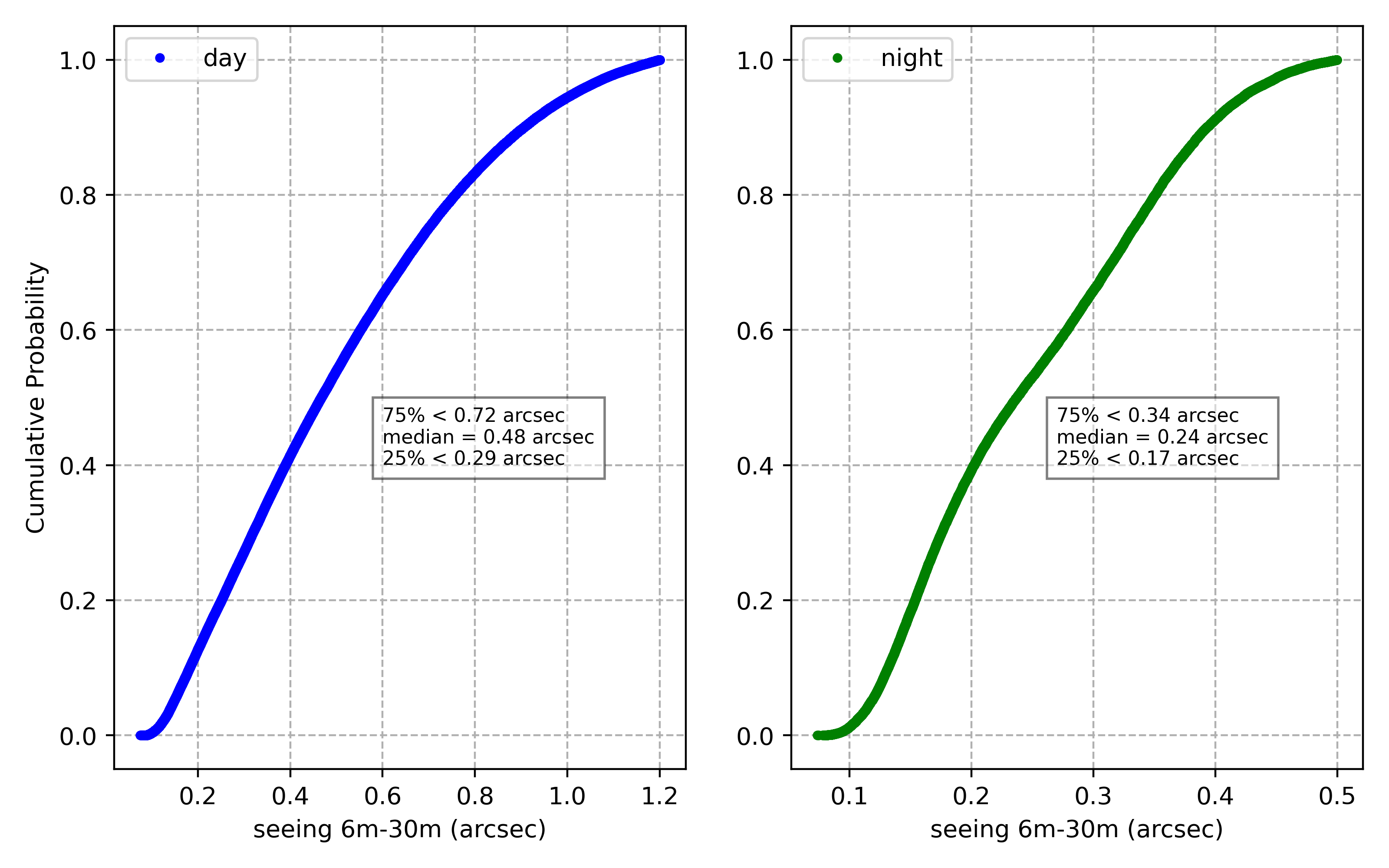}
	\vspace{-5mm} 
	\caption{The cumulative probability distribution of seeing between 6m-30m during daytime and nighttime at Muztagh-ata site.}
	\vspace{-2mm} 
	\label{fig:seeing6-30}
\end{figure}

\begin{figure}
	\centering
	\begin{minipage}{\columnwidth}
		\includegraphics[width=1.01\columnwidth]{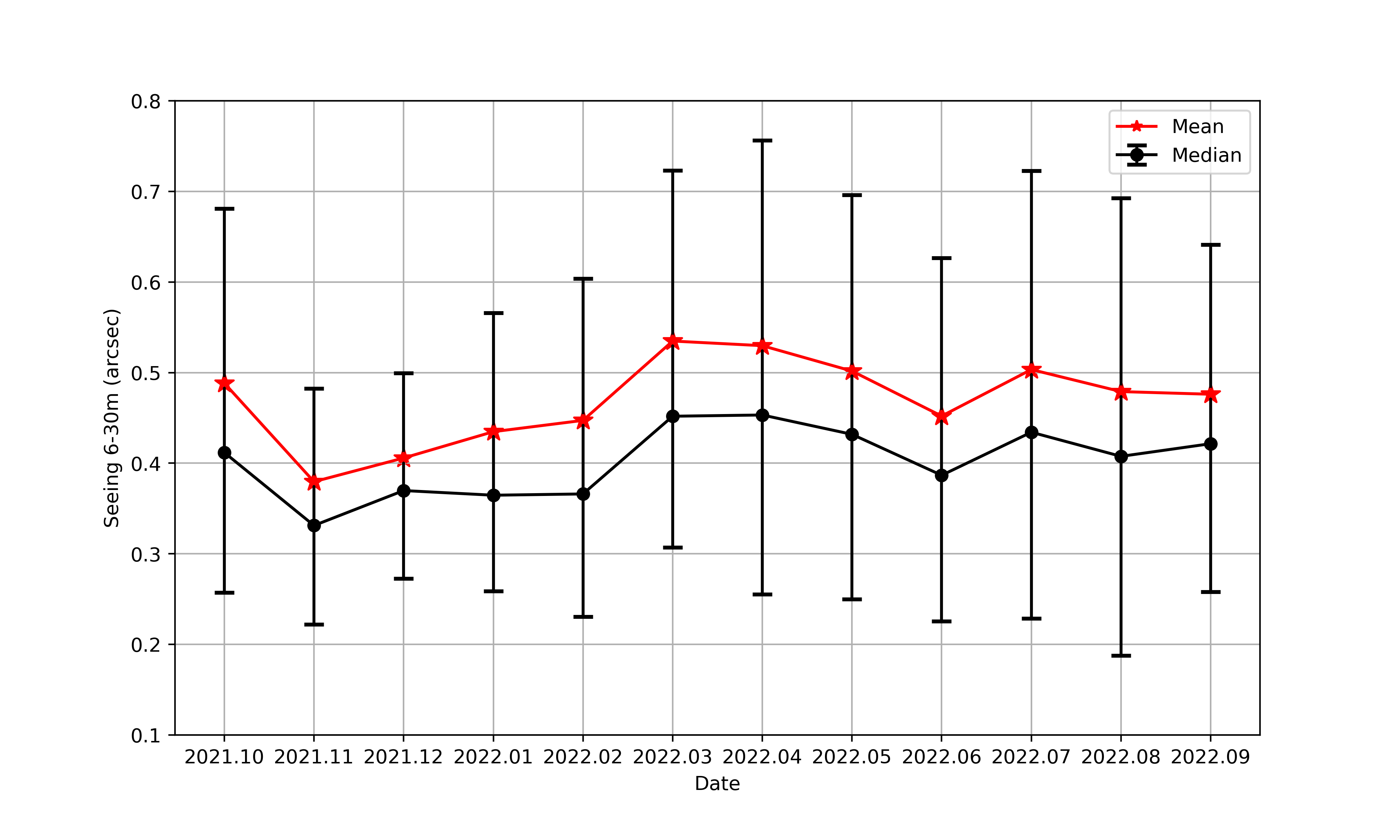}
		\vspace{-5mm} 
		\caption{The monthly evolution of seeing in the 6-30 meter range is shown. Error bars depict the 50\% confidence interval. The black line represents the monthly median of seeing, while the red line represents the monthly mean of seeing.}
		\vspace{-1mm} 
		\label{fig:fig16}
	\end{minipage}
	
	\begin{minipage}{\columnwidth}
		\includegraphics[width=1.01\columnwidth]{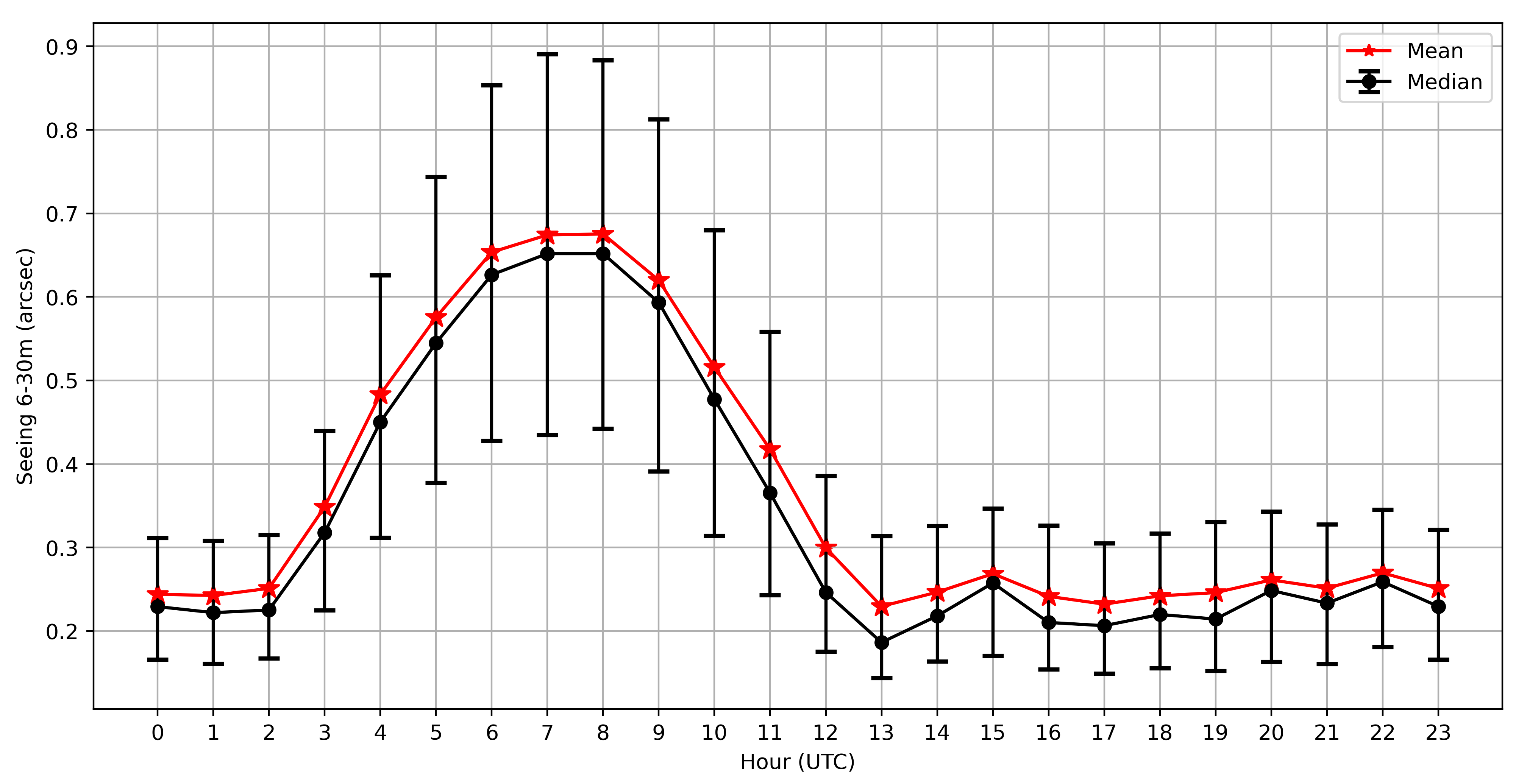}
		\vspace{-5mm} 
		\caption{The hourly variation of seeing in the 6-30 meter range is shown. Error bars depict the 50\% confidence interval. The black line represents the hourly median of seeing, while the red line represents the hourly mean of seeing.}
		\vspace{-1mm} 
		\label{fig:fig17}
	\end{minipage}
\end{figure}

Thus, we can obtain the contributions to the seeing from the five different heights.
Figure~\ref{fig:seeing6-30} shows the cumulative probability distribution of seeing at heights of 6-30 meters in the ASL during both daytime and nighttime. We find that the median daytime seeing is 0.48 arcseconds, with 75\% of the data being less than 0.72 arcseconds. The median nighttime seeing is 0.24 arcseconds, with 75\% of the data being less than 0.34 arcseconds. 

Figure ~\ref{fig:fig16} depicts the monthly evolution of quartile values of seeing within the height range of 6-30 meters, with each asterisk representing the mean seeing for each month. 
It is evident that at Muztagh-ata site, seeing is larger during the summer, and the most favorable observation period is from late autumn to early winter. This corresponds to the altitude variations observed by Xu et al ({\citeyear{xu2020b}}).
As illustrated in Figure~\ref{fig:fig17}, we demonstrate the diurnal characteristics of quartile values of seeing within the height range of 6-30 meters. Clearly, influenced by thermal radiation, near-surface seeing peaks at midday.

\subsection{ Seeing and Temperature inversion}
A stable temperature inversion layer suppresses the formation of optical turbulence. At Muztagh-ata site, rapid cooling of the ground at night may lead to the formation of a near-surface temperature inversion, where temperature increases with height. With warmer air overlaying colder air, temporarily stabilizing the lower atmosphere and suppressing turbulent mixing. Antarctic locations such as Dome A and Dome C exhibit well temperature inversion layers above the ground.
(Burton \citeyear{2010A&ABurton}; Aristidi et al. \citeyear{2015Aristidi}; Ma et al. \citeyear{2020NatureM}).

\begin{figure}
	\centering
	\includegraphics[width=\columnwidth]{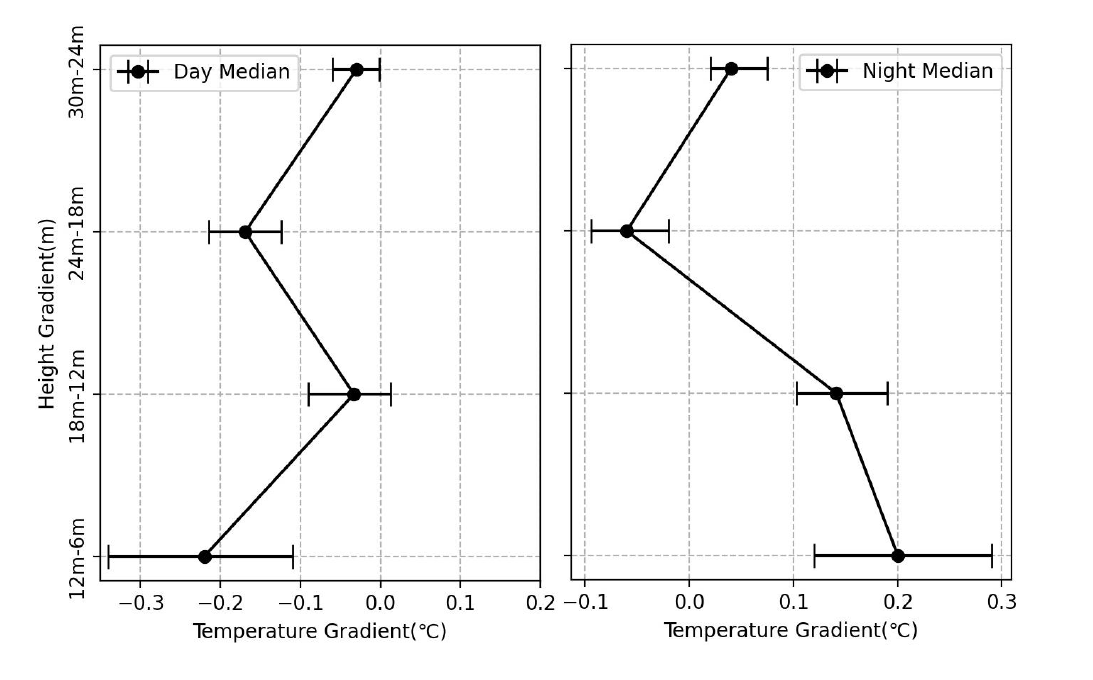}
	\vspace{-2mm} 
	\caption{The degree of temperature inversion between various layers during day and night at  Muztagh-ata site is shown. A temperature gradient greater than 0 indicates the presence of a temperature inversion. "12m-6m" represents the temperature difference between 12m and 6m heights. Error bars depict the 50\% confidence interval.}
	\vspace{-2mm} 
	\label{fig:newtemp}
\end{figure}

\begin{figure}
	\centering
	\includegraphics[width=1.02\columnwidth]{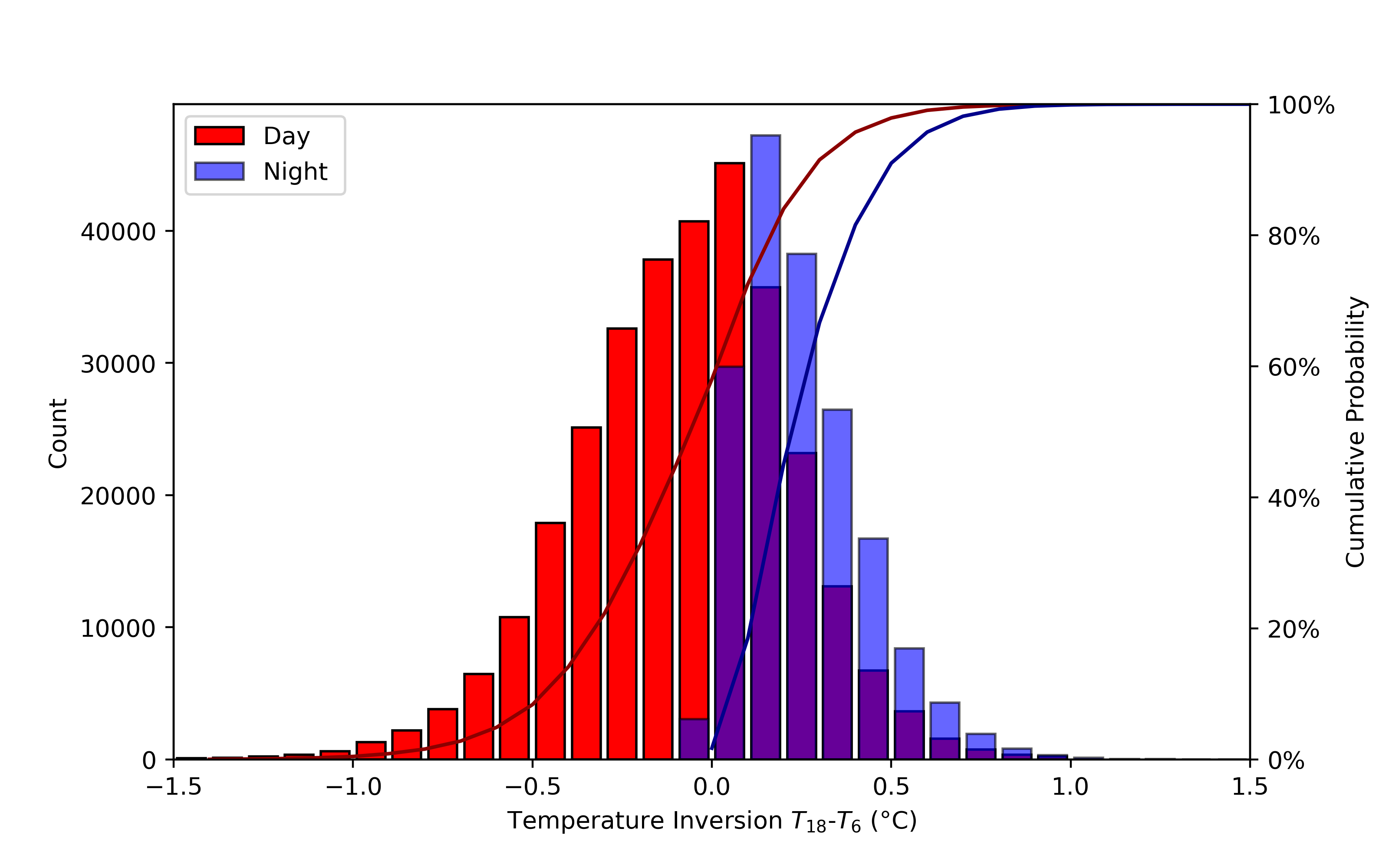}
	\vspace{-2mm} 
	\caption{Statistical analysis of temperature inversion  between 6 and 18 meters at the monitoring points in Muztagh-ata site.}
	\vspace{-2mm} 
	\label{fig:Temperature Inversion}
\end{figure}

There is a relatively stable temperature inversion layer near the surface during the night at Muztagh-ata site. 
Figure Figure~\ref{fig:newtemp} shows the temperature inversion gradients during the day and night on the 30-meter tower. A temperature gradient greater than 0 indicates the presence of a temperature inversion. During the day, temperature inversions at various heights are not significant, while at night, there are notable temperature inversions below the 18-meter height. The situation changes above 18 meters. The statistical chart in Figure~\ref{fig:Temperature Inversion} illustrates the extent of the temperature inversion between 6-18 meters. The red bars represent the statistical results of daytime temperature differences (18m-6m), while the blue bars represent the statistical results of nighttime temperature differences (18m-6m). The red and blue curves represent the probability density distributions. It is evident that  temperature inversion occurs during nearly 98\% of nighttime hours, while inversion situations are present in only 42\% of daytime hours.

\begin{figure}
	\centering
	\includegraphics[width=0.99\columnwidth]{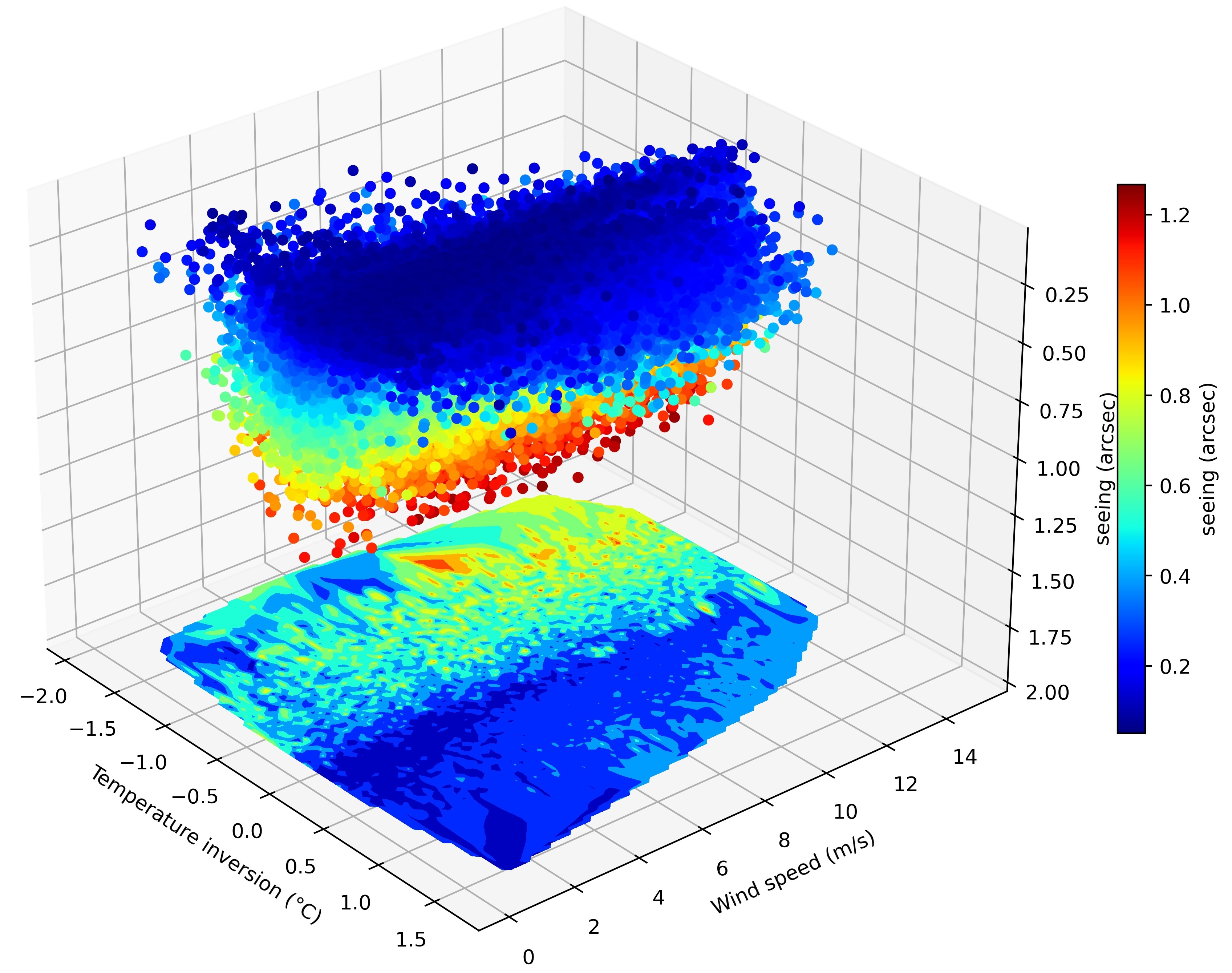}
	\vspace{-3mm} 
	\caption{Analysis of seeing (6-18m) dependence on wind speed and temperature inversion (6-18m) at Muztagh-ata site.}
	\vspace{-2mm} 
	\label{fig:fig19}
\end{figure}

Figure~\ref{fig:fig19} displays a 3D graph showing the dependency of seeing in the 6-18m layer on wind speed and temperature inversion, with the bottom layer representing the projection of seeing. Temperature inversion > 0 indicates the presence of inversion. The stratification observed in the graph suggests a significant correlation between seeing and temperature inversion, where greater temperature inversion corresponds to better seeing. Moreover, Moderate to low wind speeds can improve seeing, while excessively high wind speeds may lead to increased dynamic turbulence, thereby enhancing atmospheric instability. The pattern of change in seeing in relation to temperature inversion is consistent with the variability observed at the Delingha site (Zhu et al. \citeyear{2023MNRASzhu}).
Around $\Delta T \approx 0^\circ$C, relatively good seeing is observed, which may be related to the neutral layer formed during exchange of heat between  ground and atmosphere. This layer is characterized by minimal temperature differences between layers, high air stability, and weaker vertical airflows and turbulence. Notably, to better illustrate the relationship between temperature inversion and seeing, extreme cases with wind speeds exceeding 15 meters per second were filtered out in our analysis.


The Richardson number (Ri) quantifies the relationship between temperature gradient and wind shear effects to measure the stability of a fluid. It is widely used in analyzing atmospheric turbulence and stratification stability and is expressed as:
\begin{equation}
	Ri = \frac{g}{\Theta} \cdot \frac{d\Theta}{dz} \cdot \left(\frac{du}{dz}\right)^{-2}
\end{equation}
Where \( g \) is the gravitational acceleration, \( \Theta \) is the potential temperature, \( \frac{d\Theta}{dz} \) is the vertical gradient of potential temperature, and \( \frac{du}{dz} \) is the vertical gradient of wind speed.
When \( \text{Ri} > 0.25 \), the temperature gradient dominates, suppressing turbulence formation, and the atmosphere is typically stable. When \( 0 \leq \text{Ri} \leq 0.25 \), the atmosphere is in a neutral or weakly stable state, and turbulence may form. When \( \text{Ri} < 0 \), the atmosphere is unstable, with the velocity shear effect dominating, leading to a higher likelihood of turbulence formation (Richardson \citeyear{richardson1920supply}; Vernin \citeyear{vernin2002mechanism}).

\begin{figure}
	\centering
	\includegraphics[width=0.99\columnwidth]{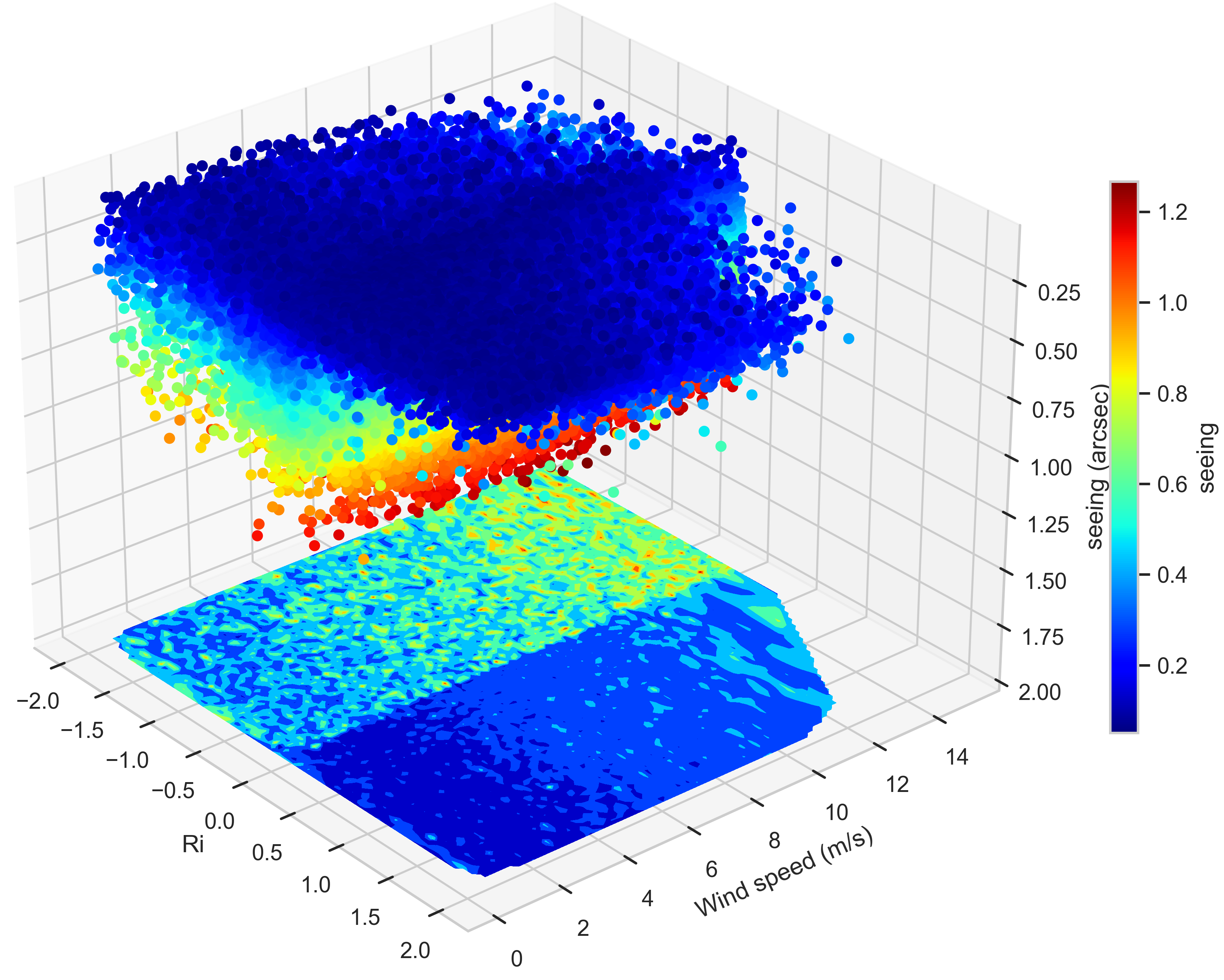}
	\vspace{-3mm} 
	\caption{Analysis of nighttime seeing(6-18m) dependence on wind speed and Ri(Richardson number) at Muztagh-ata site.}
	\vspace{-2mm} 
	\label{fig:fig20}
\end{figure}

The 3D plot in Figure~\ref{fig:fig20} and its projection show the relationship among seeing in 6-18m layer, the Richardson number (Ri), and wind speed. As observed in Figure~\ref{fig:fig20}, poorer seeing consistently occurs when Ri < 0, while seeing is generally excellent when Ri > 0, with better seeing corresponding to higher Ri values. Good seeing is associated with moderate to low wind speeds, but seeing deteriorates as wind speed increases.

\begin{figure}
	\centering
	\includegraphics[width=0.99\columnwidth]{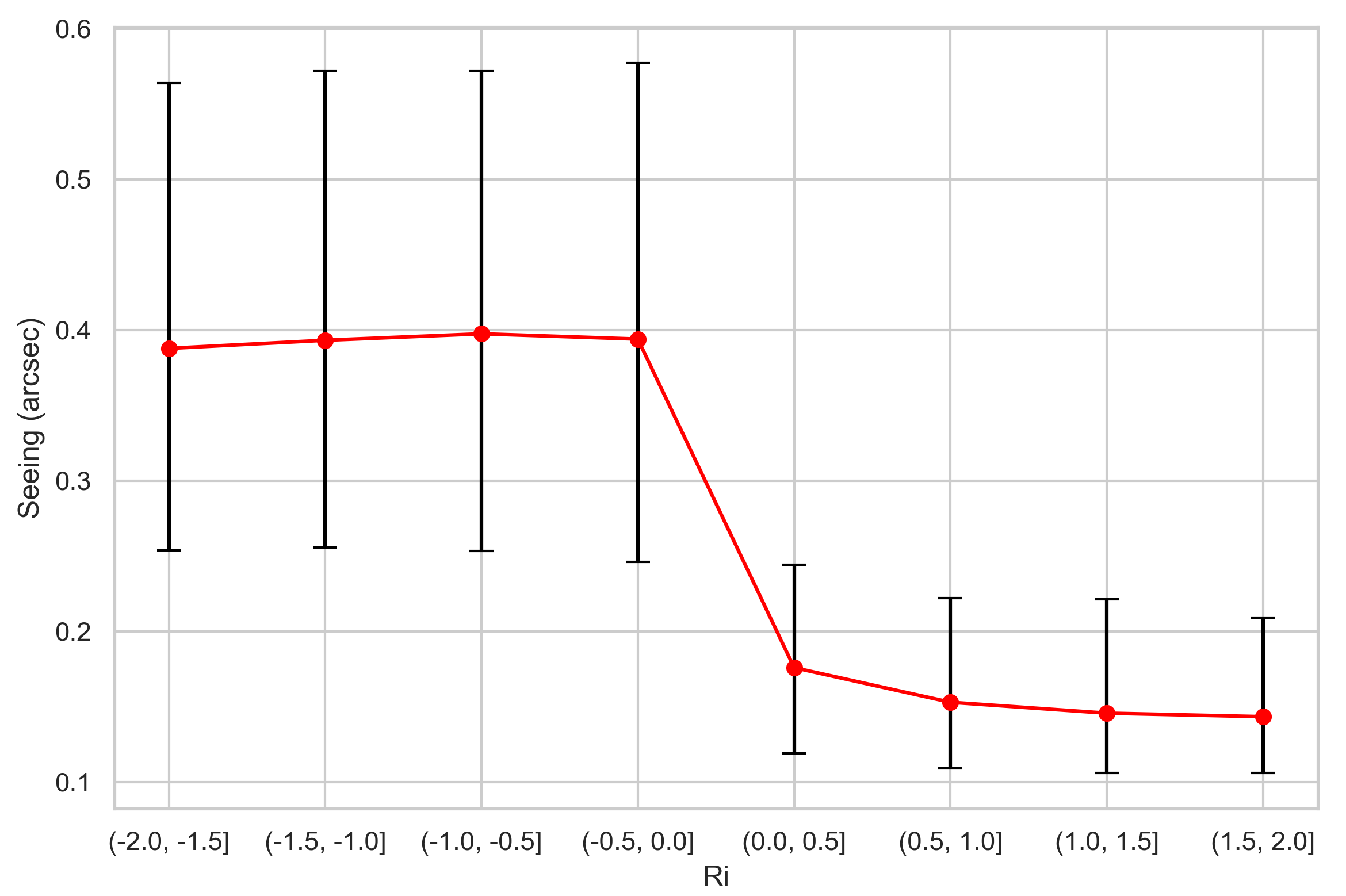}
	\vspace{-3mm} 
	\caption{The median seeing values corresponding to different ranges of Ri are shown, with the red line representing the median seeing and the error bars indicating the 50\% confidence interval.}
	\vspace{-2mm} 
	\label{fig:figRi}
\end{figure}

Figure~\ref{fig:figRi} shows the median seeing values and the 50\% confidence intervals across different ranges of Ri. When Ri < 0, the overall seeing is relatively large, with a median of around 0.4. When Ri > 0, seeing significantly decreases, indicating that a stable atmospheric state results in excellent seeing. As the Ri range increases, the median seeing value continues to decline.

\section{conclusion}
In this study, we utilized a 5-layer ultrasonic anemometer installed on a 30 m tower to investigate the characteristics of optical turbulence parameters (\(C_n^2\) and seeing \(\varepsilon\)) in the ASL at Muztagh-ata site. We explored the relationship between ASL seeing and temperature inversion.
The study period spans from October 1, 2021 to September 30, 2022. This marks the first comprehensive analysis of optical turbulence in the ASL measured on-site at the site.
The main conclusions are as follows:

(1) At Muztagh-ata site, the optical turbulence intensity \(C_n^2\) in the ASL ranges from \(10^{-16}\, \mathrm{m}^{-2/3}\) to \(10^{-13}\, \mathrm{m}^{-2/3}\) during the day and from \(10^{-16}\, \mathrm{m}^{-2/3}\) to \(10^{-14}\, \mathrm{m}^{-2/3}\) at night. Notably, \(C_n^2\) is significantly higher during the day than at night. We analyzed the vertical variation of \(C_n^2\) in the ASL and found an exponential decrease with altitude. The daytime height dependence is mainly around \(h^{-0.82}\), while the nighttime height dependence is around \(h^{-0.48}\). 

(2) We obtained the distribution of seeing across five height layers in the ASL. The seeing between layers decreases with increasing height, with the median seeing from 6 to 30 meters during the day being 0.48 arcseconds, and the median seeing at night being 0.24 arcseconds. The intensity of dynamic optical turbulence during the day is significantly greater than at night.

(3) We analyzed the dependency of seeing, temperature inversion and wind speed. Larger temperature inversions and moderate to low wind speeds correspond to better seeing. Seeing is influenced by both wind speed and temperature inversion. We used the Richardson number(Ri) to describe the dependency of seeing. Poor seeing generally occurs when \( Ri < 0 \), while \(\text{Ri} \geq 0\) corresponds to overall excellent seeing. The greater the \( Ri \), the better the seeing, and the results meet expectations.

\section{acknowledgment}
We sincerely thank the reviewer and the editor for their helpful suggestions, which have improved this paper a lot. This work is supported by the Chinese Academy of Science(CAS) “Light of West China” Program(Grant No.2022 XBONXZ 014), the National Natural Science Foundation of China (Grant No: U2031209), Tianshan Talent Training Program (Grant No. 2023TSYCLJ0053), Natural Science Foundation of Xinjiang Uygur Autonomous Region (Grant No. 2023D01A13).

\section{data availbleity}
Due to specific national policy restrictions, the raw data of this study cannot be made publicly available. However, detailed data descriptions can be provided upon reasonable request. Please contact the corresponding author for more information.






\bibliographystyle{mnras}


\bsp	
\label{lastpage}

\end{document}